\documentclass[10pt,aps,prl,reprint,twocolumn, nofootinbib,aps,superscriptaddress]{revtex4-1}

\usepackage{amssymb,amsmath}
\usepackage{graphicx}
\usepackage{color}
\usepackage{listings}
\usepackage{float}
\usepackage{wrapfig}
\usepackage{physics}
\usepackage{amsmath}
\usepackage{gensymb}
\usepackage{soul}
\usepackage{caption}
\usepackage{subcaption}

\usepackage{upgreek}
\usepackage{enumerate} 
\usepackage[linkcolor = blue, citecolor = blue, urlcolor = blue, colorlinks = true]{hyperref}

\usepackage[version=3]{mhchem}
\usepackage{commath,amssymb}
\usepackage{ushort}
\usepackage{multirow}
\usepackage[usenames,dvipsnames]{xcolor}
\usepackage{cleveref}
\usepackage{epstopdf}
\usepackage[exponent-product = \cdot]{siunitx}

\crefname{equation}{Eq.}{Eqs.}
\crefname{figure}{Fig.}{Figs.}
\crefrangeformat{equation}{Eqs.~#3(#1)#4--#5(#2)#6}

\hypersetup{
    colorlinks=true,
    linkcolor=blue,
    citecolor=red}

\begin{document}

\title{Diffusion-based height analysis reveals robust microswimmer-wall separation}

\author{Stefania Ketzetzi}
\affiliation{Soft Matter Physics, Huygens-Kamerlingh Onnes Laboratory, Leiden University, P.O. Box 9504, 2300 RA Leiden, The Netherlands}
\author{Joost de Graaf}
\affiliation{Institute for Theoretical Physics, Center for Extreme Matter and Emergent Phenomena, Utrecht University, Princetonplein 5, 3584 CC Utrecht, The Netherlands}
\author{Daniela J. Kraft}
\affiliation{Soft Matter Physics, Huygens-Kamerlingh Onnes Laboratory, Leiden University, P.O. Box 9504, 2300 RA Leiden, The Netherlands}

\date{\today}

\begin{abstract}
Microswimmers typically move near walls, which can strongly influence their motion. However, direct experimental measurements of swimmer-wall separation remain elusive to date. Here, we determine this separation for model catalytic microswimmers from the height dependence of the passive component of their mean-squared displacement. We find that swimmers exhibit ``ypsotaxis'', a tendency to assume a fixed height above the wall for a range of salt concentrations, swimmer surface charges, and swimmer sizes. Our findings indicate that ypsotaxis is activity-induced, posing restrictions on future modeling of their still debated propulsion mechanism. 
\end{abstract}

\maketitle

Confining surfaces, such as planar walls, have a far-reaching impact in the microswimmer world, often ensuring microswimmer function and survival~\cite{Elgeti2015}. Encounters with surfaces give rise to accumulation, as seen for sperm~\cite{Rothschild1963}, algae~\cite{Kantsler2013} and bacteria~\cite{Berke2008}, and enable the formation of bacterial biofilms that facilitate their spreading, cooperation, and capture of nutrients~\cite{Flemming2016, Berne2016, Flemming2019}. Moreover, surfaces can significantly modify swimming trajectories; \textit{e.g.}, bacteria often exhibit circular motion with direction controlled by the boundary condition~\cite{DiLuzio2005, Lauga2006, Lemelle2010, Lopez2014}, in stark contrast to their run-and-tumble motion in bulk. 

Striking surface effects are not only found in biological systems, but are also present for synthetic microswimmers~\cite{Das2015,  Simmchen2015, Brown2016, Wei2018sub, Holterhoff2018, Ketzetzi2020, Heidari2020}. Model catalytic colloidal swimmers exhibit autonomous directed motion due to self-generated chemical gradients~\cite{Ebbens2010}. Recently, neighboring walls were shown to significantly alter the magnitude of their swim speeds~\cite{Wei2018sub, Holterhoff2018, Ketzetzi2020, Heidari2020}. This revealed that walls play a far greater than previously expected role on self-propulsion, providing a path towards resolving seemingly conflicting experimental observations. For example, speed differences under similar conditions may stem from the phoretic interplay between the hydrodynamic boundary condition on the wall and the out-of-equilibrium chemical species generated by the swimmer~\cite{Ketzetzi2020}. Current models predict a wide range of behaviors close to walls, including hovering, sliding, forward and/or backward propulsion~\cite{Das2015, uspal2016guiding, Popescu2009, Crowdy2013, Chiang2014, Uspal2015,Uspal2015rheotaxis, Ibrahim2015, Mozaffari2016, Shen2018, Ebbens2012, popescu2018effective, spagnolie2012hydrodynamics, lintuvuori2016hydrodynamic, kuron2019}. This diversity is partly due to the complexity of and uncertainties in the propulsion mechanism, and partly due to the hydrodynamic and numerous phoretic couplings that wall proximity can introduce. Thus, quantitative insight into swimmer-wall separation is pivotal to pinpointing missing details of the propulsion mechanism, and in turn tailoring swimming behaviors,~\textit{e.g.}, for guiding microswimmers in complex environments.

To date, no reported experiment has directly measured swimmer-wall separations. However, based on qualitative observations, separations are anticipated to be smaller than the swimmer size~\cite{Simmchen2015, Brown2014}, even as small as a few tens of nm~\cite{Takagi2013,Takagi2014}. Such separations cannot be directly resolved by standard optical microscopy~\cite{Takagi2013}, which is why holographic microscopy has been proposed~\cite{Holterhoff2018}, as it yields three-dimensional positions of spherical particles with high precision~\cite{LeeGrier2007}. However, fitting holograms of spheres half-coated with a metal is computationally expensive, especially when studying dynamics, since discrete dipole approximations have to be employed in the numerical calculations to obtain their positions~\cite{WangManoharan2014}. Furthermore, inhomogeneities in the metal coating introduce additional fit parameters and uncertainties in determining particle positions. Another way to measure small particle-wall separations is Total Internal Reflection Microscopy (TIRM), which yields separations from the scattering of evanescent waves off of particles close to a wall~\cite{BrownTIRM1990}. Here too, the asymmetric coating interferes with interpreting the result and obtaining accurate measurements. Hence, a novel measuring approach is needed.

In this Letter, we present a facile and straightforward method for obtaining microswimmer-wall separations \textit{in situ}. We determine the translational diffusion coefficient of the swimmer from mean-squared displacement curves, and obtain the height from its theoretically predicted dependence on swimmer-wall separation. Our method can be applied to most synthetic microswimmers, and may be extended to a range of swimming microorganisms, moving parallel to walls. We applied it here to catalytically propelled model microswimmers. Besides the fuel concentration, we systematically varied additional parameters known to affect self-propulsion, as well as particle-wall separations in passive systems: the salt concentration in solution, swimmer size, and swimmer zeta potential. We were thereby able to gain unprecedented insights into their effect and the presence of a wall on the swimming behavior. 

\begin{figure*}[!ht]
\centering
  \includegraphics[width=1\linewidth]{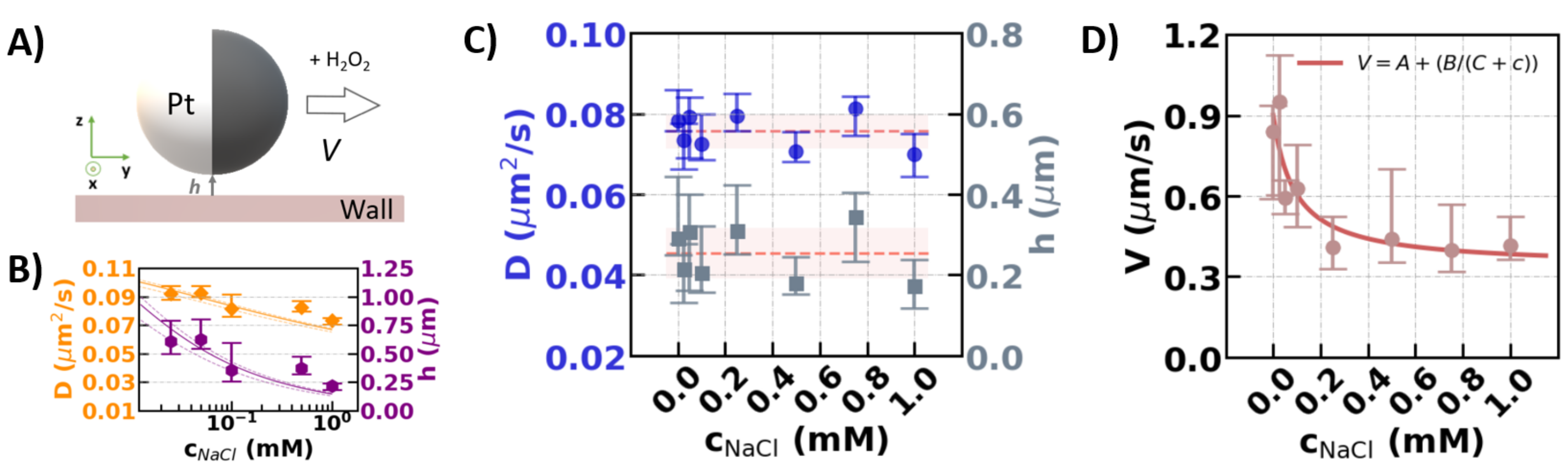}
\caption{\label{fig:1}\textbf{Salt-dependent motion above a planar wall:} A) Schematic of the experiment. We obtain the swimmer-wall separation, $h$, from the measured translational diffusion coefficient, $D$, of the swimmer and its theoretically predicted dependence on wall separation. B), C), and D): Effect of salt concentration $c_{\mathrm{NaCl}}$ on the motion of 2.77~$\pm$~0.08~$\mu$m colloids with 4.4~$\pm$~0.2~nm Pt. All reported values are medians and error bars denote 1$^{\mathrm{st}}$ quartiles. B) Diffusion coefficient (orange diamonds) and separation (purple hexagons) in the Brownian state in water with $c_{\mathrm{NaCl}}$. Lines show theoretical predictions based on balancing electrostatics and gravity. C) Diffusion coefficient (circles) and separation (squares) in the active state in aqueous 10\% H$_{2}$O$_{2}$ with $c_{\mathrm{NaCl}}$. Dotted lines indicate mean values. D) Speed decrease in 10\% H$_{2}$O$_{2}$ with $c_{\mathrm{NaCl}}$. Solid line is a least-squares fit with $V = A + (B/(C+c_{\mathrm{NaCl}}))$, where $A$ is 0.35~$\pm$~0.09~$\mu$m/s the remaining speed in high salt, $B$ a prefactor, and $C$ is 0.09~$\pm$~0.07~mM the ion concentration already present in solution, following from ionic diffusioosmosis along the wall.}
\end{figure*}

We obtained swimmer-wall separations from experimental measurements of the separation-dependent translational diffusion coefficient, $D$, of the swimmers. $D$ as well as propulsion speeds $V$ were extracted from mean square displacements (MSDs) following Ref.~\cite{Howse2007,Bechinger2016}. That is, we fitted the short-time regime ($\Delta\mathrm{t} \ll \tau_R$) of the MSDs with $\Delta r^{2} = 4 D \Delta t + V^{2} \Delta t^{2}$~\cite{Howse2007}; $\tau_{\mathrm{R}}$ is the rotational diffusion time, $\tau_R = 1/D_{\mathrm{R,bulk}}$,
with $D_{\mathrm{R,bulk}} = \frac{k_{\mathrm{B}} T}{8\pi\eta R^3}$ the bulk rotational diffusion coefficient, $R$ the radius, $\eta$ the viscosity, $k_{B}$ the Boltzmann constant, and $T$ the absolute temperature. The first term corresponds to the passive diffusion contribution that is usually obscured by the activity-induced, short-time ballistic behavior~\cite{Bechinger2016}, but may be obtained with sufficient statistics. Reliable measurements require frame-rate adjustment, such that the regime where both diffusion and activity contribute to the MSD can be resolved. See the Supplemental Information (SI)~\cite{supp}, which additionally includes Ref.~\cite{henry1931electrokinetic,degraaf2015raspberry,BocquetAngle2008,eloul2020,zhou2018photochemically}, for details on tracking, MSD calculation~\cite{Trackpy}, as well as a discussion on the consistent short-time ($\Delta t \ll \tau{_R}$) expansion of the MSD~\cite{Howse2007, Palacci2010, Bechinger2016}. 
To calculate the separation, $h$, between the particle and wall, see also Figure~\ref{fig:1}A, we first consider the ratio $d = D/D_{\mathrm{bulk}}$, with $D_{\mathrm{bulk}} = \frac{k_B T}{ 6 \pi \eta R }$ the bulk diffusion constant. For $d$~$\gtrsim$~0.6, the well-known prediction by Fax{\'e}n~\cite{faxen1921einwirkung,oseen1927neuere,Sharma2010Faxen, Ha2013Faxen}: $d(h) = 1 - \frac{9}{16} \gamma + \frac{1}{8} \gamma^{3} - \frac{45}{256} \gamma^{4} - \frac{1}{16} \gamma^{5} $, with $\gamma = R/(h+R)$, can be used to extract $h$. For $d$~$\lesssim$~0.4 a lubrication theory result,~$d(h) = -\frac{1}{ \frac{8}{15} \log\left( \frac{h}{R} \right) - 0.9588}$, is more appropriate~\cite{goldman1967slow,oneill1964slow}. In the intermediate (0.4 $\lesssim d \lesssim$ 0.6) regime, applicable to most of our experiments, we interpolate the combined numerical data by O’Neill~\cite{oneill1964slow} and Kezirian~\cite{kezirian1992hydrodynamics}, see SI Section II-A~\cite{supp}. The $d(h)$ relation that follows is provided as supplement to this publication. Here, we fitted for $h$ using the interpolated expression.

In all experiments, we used 3-(trimethoxysilyl)propyl methacrylate (TPM) monodisperse colloids~\cite{Wel2017TPM} half-coated with a thin Pt layer ($\approx$ 4.5 $\pm$ 0.2 nm) at dilute concentration ($\approx~10^{-7}$~v/v). In water, colloids exhibited passive Brownian motion, while dispersion in 10\% H$_{2}$O$_{2}$ rendered them active through a catalytic process. Colloids quickly reached the lower glass wall and continued to move adjacent to it, while their motion was recorded with an inverted Nikon Eclipse Ti microscope through a 60x oil objective (NA=1.4). Swimming experiments were recorded for 30 s at 19~fps, see SI, Section I-C~\cite{supp}.

\begin{figure*}[!ht]
\centering
  \includegraphics[width=1.0\linewidth]{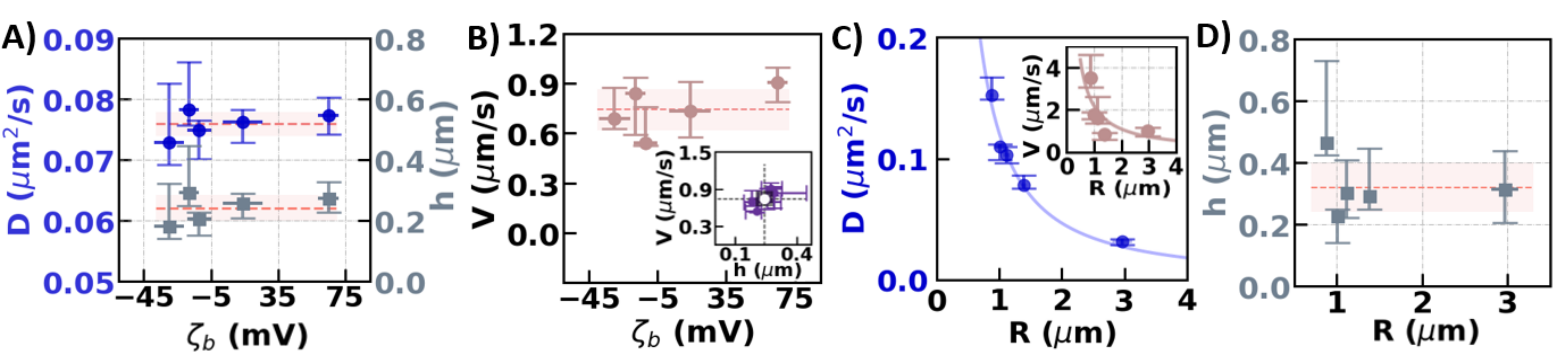}
\caption{\label{fig:2}\textbf{Swimmer base zeta potential and size dependence of propulsion above a planar wall.} All reported values are medians, error bars denote 1$^\mathrm{st}$ quartiles, and dotted lines represent mean values. All experiments were performed in aqueous 10\% H$_2$O$_2$. A), B): The base zeta potential, $\zeta_{b}$, of TPM colloids with diameters between 2.70~$\pm$~0.06 and 2.77~$\pm$~0.08~$\mu m$, and Pt coating thicknesses $\approx$ 4.4 $\pm$ 0.2 nm, was varied through surface functionalization. A) Diffusion coefficient (circles) and separation (squares) with $\zeta_{b}$. B) Speed for the same $\zeta_{b}$ range as in (A). Inset shows speed with separation (purple circles), with the write circle marking the intersection of mean values, and the rectangle around it denoting standard deviations. C), D): Variation of the radius, $R$, of active TPM spheres with similar $\zeta_b$ and Pt (coating thickness $\approx$ 4.5 $\pm$ 0.2 nm) affects C) the diffusion coefficient and D) swimmer-wall separation. Inset in (C) shows swim speed with $R$. Solid lines in (C) are fits with $a/R^b$ with $a$~0.120~$\pm$~0.004~$\mu$m$^3$/s and b~1.3~$\pm$~0.2 (main), and the expected $a/R$~\cite{Ebbens2012} with $a =$~2.2~$\pm$~0.4~$\mu$m$^2$/s (inset). Dotted line in (D) shows mean separation (0.32~$\pm$~0.08~$\mu$m).} 
\end{figure*}

To demonstrate the effectiveness of our method, we first carried out control experiments in deionized water and in water at pH 3.3, equivalent to the pH in the swimming experiments, at 5~fps. In these cases, $D$ was acquired from fitting MSDs with $\Delta r^2=4 D \Delta t$. Figure~\ref{fig:1}B shows that the extracted separation corresponds well to a theoretical prediction based on a balance of electrostatic repulsion and gravity~\cite{flicker1993measuring, Rashidi2017}, see the SI Section II-B~\cite{supp}. That is, we recovered the expected decrease in separation with increasing salt concentration: salt increases the solution's ionic strength, thereby effectively screening the charge on the particle and wall. This reduces the Debye length,~\textit{i.e.}, the distance over which surface charges act, bringing the colloids closer to the wall. To verify our method further, we compared separations resulting from our diffusion coefficient-based method to those directly measured with DIHM, for uncoated silica spheres with well-known size and refractive index~\cite{KetzetziVerweij2020}. We found good agreement between the two methods, which confirmed that we indeed recovered colloid-wall separations, using a computed rather than a measured value of D$_{\mathrm{bulk}}$.

Having established the validity of our method, we employed it to our catalytic microswimmers. First, we studied the effect of salt concentration in solution. For these experiments, we used TPM spheres of 2.77~$\pm$~0.08~$\mu$m diameter half-coated with 4.4~$\pm$~0.2~nm Pt. Surprisingly, in active systems we found a behavior completely unlike that of passive systems in Figure~\ref{fig:1}B. For the same particles and salt concentration range, $D$ and $h$ remain constant within measurement precision, see Figure~\ref{fig:1}C. Particles propel themselves parallel to the wall at constant separations of 0.25~$\pm$~0.06~$\mu$m.

At the same time, we found a decrease in speed with increasing salt concentration, see Figure~\ref{fig:1}D, where the line represents the least-squares fit with $V = A + (B/(C+c))$. This expression follows from a salt-gradient based contribution to the observed speed~\cite{Brown2014}, with $A$ the remaining speed in the limit of high salt, $B$ a prefactor, and $C$ the ion concentration already present in the medium. From the fit we find the reasonable numbers 0.35~$\pm$~0.09~$\mu$m/s and 0.09~$\pm$~0.07~mM, for $A$ and $C$, respectively. The fitted C value agrees reasonably well with the background ion concentration (0.008 mM) we obtained from electrical conductivity measurements~\cite{Coury1999} for 10\% H$_2$O$_2$ (2.7~$\mu$S/cm, Ilium technology, Model 2100 Conductivity Meter) assuming hydrogen ions as the dominant ion species. We return to this salt gradient in the discussion. 

Second, we explored the effect of colloid zeta potential, $\zeta$, the electric potential at the colloid's surface. We used 2.70~$\pm$~0.06 and 2.77~$\pm$~0.08~$\mu$m diameter colloids with different surface functionalizations~\cite{Wel2017} and thus different $\zeta$. The reported $\zeta$ correspond to those of the parent colloids, see SI Sections I-A and I-B~\cite{supp} for characterization, before adding the Pt-coating. We therefore use the adjective ``base'' and a subscript ``b'',~\textit{i.e}, $\zeta_{b}$, to indicate that we know only the zeta potential of the uncoated colloid, and not that of the swimmer. We note that passive colloids with $\zeta_{b} > -12$~mV were typically stuck on the negatively charged wall, see also SI Section~I-D~\cite{supp}.

However, for the active system, we found that wall separation remained unaffected for the entire (wide) range of $\zeta_{b}$ under study, see Figure~\ref{fig:2}A. In all cases, particles moved at 0.24~$\pm$~0.04~$\mu$m from the wall, which matches the separations measured for different salt concentrations. Unexpectedly, as we will return to, the colloids self-propelled not only at a constant $h$ when varying $\zeta_{b}$, but also at quantitatively comparable speeds $V$, see Figure~\ref{fig:2}B. We can indeed collapse the data by plotting $V$ as a function of $h$, see inset of Figure~\ref{fig:2}B, further demonstrating that $\zeta_{b}$ does not affect the swimming behavior. 
We note that the direction of motion was away from the Pt cap both for positive and negative $\zeta_{b}$.

\begin{figure*}[!ht]
\centering
  \includegraphics[width=0.9\linewidth]{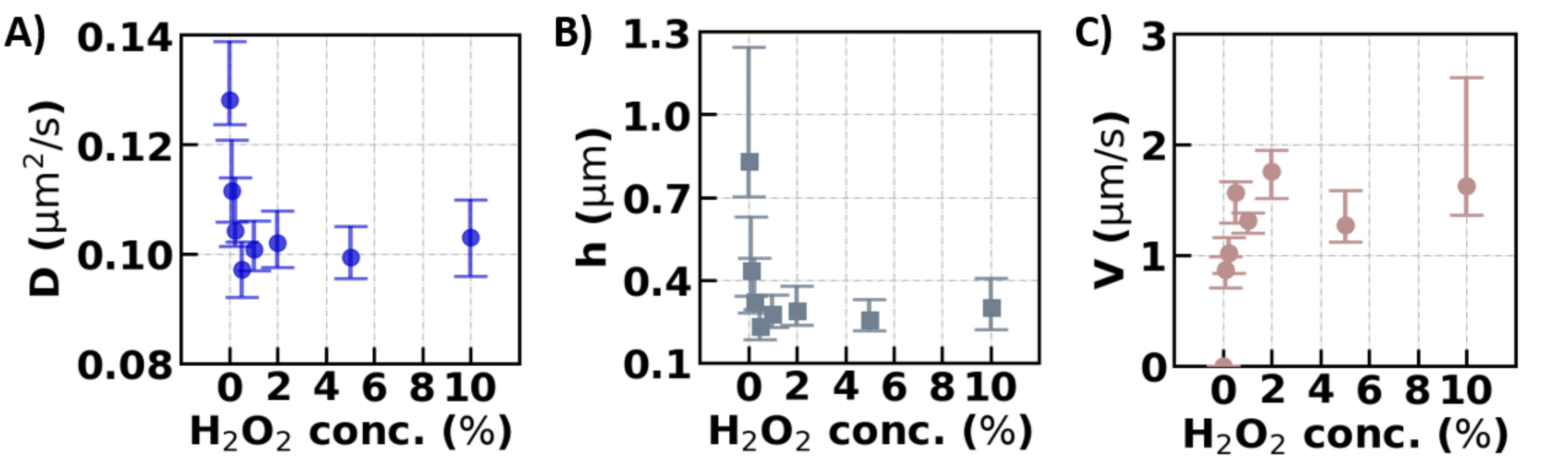}
\caption{\label{fig:3}\textbf{Activity-induced ypsotaxis}.  A) Diffusion coefficient $D$ with increasing H$_{2}$O$_{2}$ (fuel) concentration. B) Equivalent swimmer-wall separation $h$ for the same fuel concentration range. C) Swim speed $V$ for the same fuel concentration range as in (A,B). All experiments were performed using 2.23 $\pm$ 0.11 $\mu$m diameter TPM spheres with 4.5 $\pm$ 0.2 nm Pt. All reported values are medians and error bars denote 1$^\mathrm{st}$ quartiles.}
\end{figure*}

Third, we focused on swimmer size, another parameter known to affect swim speeds~\cite{Ebbens2012}. We performed experiments using TPM spheres with a wide range of radii, but with similar Pt coating thicknesses and $\zeta_b$, see SI Section I-A for characterization~\cite{supp}. We found that diffusion coefficient decreases with swimmer size, see Figure~\ref{fig:2}C, where the solid line represents the least-squares fit with $D=a/R^{b}$ ($a = 0.120 \pm 0.004~\mu$m$^3$/s, $b = 1.3 \pm 0.2$). Inset shows the measured swim speeds together with a fit of the expected scaling $V = a/R$~\cite{Ebbens2012} ($a =2.2 \pm 0.4~\mu$m$^2$/s). Strikingly, swimmer-wall separation remained relatively constant with $R$, see Figure~\ref{fig:2}D; the dashed line shows the mean separation of 0.32~$\pm$~0.08~$\mu$m. 

The above experiments reveal that our swimmers exhibit ``ypsotaxis'': a tendency to assume a specific height for a wide range of parameters. Remarkably, the height appears independent of salt concentration, $\zeta_b$, and even size, running not only counter to our intuition for passive systems but also to features of common self-propulsion mechanisms. For our Pt-coated swimmers, this robust separation distance was found to be 0.27~$\pm$~0.11~$\mu$m on average, in line with the observation that micron-sized catalytic swimmers do not self-propel over steps of a few hundred nanometers~\cite{Simmchen2015}. Such a height is further consistent with wall-dependent speeds~\cite{Holterhoff2018, Wei2018sub, Ketzetzi2020, Heidari2020}, for which wall separation must not substantially exceed the swimmer size to ensure strong osmotic coupling~\cite{Das2015, Ibrahim2015, uspal2016guiding}. The wide range of swimmer sizes employed here showed that buoyancy is not the prime contributor to ypsotaxis. This is further underpinned by our observation of swimmers moving along the top wall, upon inversion of the sample holders, for a period of time. We hypothesize that ypsotaxis is instead primarily caused by phoretic and osmotic flows, \textit{i.e.} it is activity-driven.

To test for this, we performed experiments using 2.23 $\pm$ 0.11 $\mu$m diameter TPM spheres with 4.5 $\pm$ 0.2 nm Pt for various H$_2$O$_2$ concentrations, and hence degrees of activity. Indeed, we found that diffusion coefficient and thus swimmer-wall separation not only decreases rapidly with increasing fuel concentration from the Brownian state (0\% H$_2$O$_2$), see Figure~\ref{fig:3}A and B, respectively, but also plateaus beyond 0.25\% H$_{2}$O$_{2}$. Similarly, the speed also increases sharply and then plateaus above 0.25\% H$_2$O$_2$, see Figure~\ref{fig:3}C. These observations imply that the constant separation distance is induced by the activity, thereby confirming our hypothesis on the origin of ypsotaxis. We argue that said origin also causes the active alignment of catalytic swimmers with respect to the wall~\cite{Das2015, Simmchen2015, Ebbens2011, supp}, see discussion in SI Section II-D~\cite{supp}.

Our results provide new insights into the debated nature of the propulsion mechanisms~\cite{Ebbens2014}. Current thinking favors self-electrophoresis~\cite{Brown2014, Ibrahim2017},~\textit{i.e.}, motion generated \textit{via} self-generated ionic currents, as simple salts are known to greatly decrease propulsion speeds. The lack of speed variation with $\zeta_{b}$, however, is not commensurate with this, or other ion-based propulsion mechanisms, typically scaling with $\zeta$ or $\zeta^2$, see~\cite{Brown2017}. A possible explanation is that a different $\zeta$ at the Pt cap dominates the swimmer's behavior.

However, speed variation with salt --- typically indicative of a change in activity --- is not readily reconciled with a constant $h$ which is also activity-driven, even if the cap's $\zeta$ dominates. Drawing upon our previous work~\cite{Ketzetzi2020}, we provide an alternative wall-centric explanation: Suppose that the swimmer's bulk speed is unaffected by adding salt. The swimmer's effective near-wall speed may still vary, provided salt impacts the osmotic counterflow induced by the swimmer-generated chemical species interacting with the wall~\cite{Ketzetzi2020}. Our fit in Figure~\ref{fig:1}D reveals that the osmotic contribution to the speed bears the hallmarks of ionic diffusion~\cite{Anderson1989}. This requires a net-neutral gradient of ions with different electric mobilities to be involved, often referred to as a salt gradient. This salt gradient might originate from the chemical dissociation reactions in the long-range H$_2$O$_2$ gradient with the wall~\cite{Brown2017}, stemming from fuel consumption at the Pt cap. This model would have the right features to show an ionic diffusioosmosis along the wall, see SI Section II-C~\cite{supp}.

In summary, we established a novel method for measuring microswimmer-wall separations utilizing the height dependence of the diffusive component of their mean-squared displacement. We found that catalytic model microswimmers propel at roughly fixed heights of few hundred nanometers from planar walls. Our work further showed that nearby walls could be dominant factors in controlling swim speeds,~\textit{i.e.}, ion-induced flow may only play a role at the wall, and not at the swimmer surface. This would necessitate a paradigm shift in modeling experimental observations and in identifying the still missing details of their propulsion mechanism. Our method can be readily applied to other types of spherical microswimmers moving parallel to walls, and may be extended to different swimmer shapes as well. We are confident that further application of our method will provide novel insights on the impact of confining surfaces in the microswimmer world, and in turn facilitate predicting swimming behaviors in complex environments.

\acknowledgements

We gratefully acknowledge Rachel Doherty for providing TPM colloids and for discussions on colloid functionalizations. We thank Ruben Verweij and Nikos Oikonomeas for useful discussions on holographic microscopy and Aidan Brown for discussions on the propulsion mechanism and for pointing out a relevant passage in the literature. J.d.G. thanks NWO for funding through Start-Up Grant 740.018.013 and through association with the EU-FET project NANOPHLOW (766972) within Horizon 2020. D.J.K. gratefully acknowledges funding from the European Research Council (ERC) under the European Union's Horizon 2020 research and innovation program (grant agreement no. 758383). 

\bibliography{bbl}
\bibliographystyle{unsrtnat}

\end{document}